{}
{}

\newif\ifniklas\niklastrue
\newif\ifarxiv\arxivtrue

\arxivtrue     

\ifarxiv
\documentclass[12pt,a4paper]{article}

\usepackage{ifpdf}
\ifniklas\ifpdf\else
\PassOptionsToPackage{hypertex}{hyperref}
\fi\fi

\usepackage{amsmath,amssymb}
\usepackage[bookmarks=true]{hyperref}
\usepackage[nosort]{cite}
\usepackage[bulletsep]{collref}

\providecommand{\hypersetup}[1]{}

\providecommand{\pdfbookmark}[3][]{}

\hypersetup{pdftitle={Boosting Nearest-Neighbour to Long-Range Integrable Spin Chains}}
\hypersetup{pdfauthor={Till Bargheer, Niklas Beisert, Florian Loebbert}}
\hypersetup{pdfsubject={PACS (2010): 02.30.Ik, 75.10.Pq, 11.25.Tq}}
\hypersetup{pdfkeywords={Integrable Spin Chains, Long-Range Interactions, Boost Operators}}
\hypersetup{plainpages=false}
\hypersetup{pdfstartview=FitH}
\hypersetup{pdfpagemode=UseOutlines}
\hypersetup{bookmarksnumbered=true}
\hypersetup{bookmarksopen=true}
\hypersetup{colorlinks=false}
\hypersetup{citebordercolor={.6 .9 .6}}
\hypersetup{urlbordercolor={.7 .8 1}}
\hypersetup{linkbordercolor={1 .7 .7}}
\hypersetup{pdfborder={0 0 .5}}

\usepackage[a4paper,text={450pt,650pt},centering]{geometry}

\usepackage[font=small,labelfont=bf,width=0.85\textwidth]{caption}

\expandafter\def\expandafter\bfseries\expandafter{\bfseries\ifmmode\else\boldmath\fi}
\expandafter\def\expandafter\mdseries\expandafter{\mdseries\ifmmode\else\unboldmath\fi}
\expandafter\def\expandafter\normalfont\expandafter{\normalfont\ifmmode\else\unboldmath\fi}

\else

\documentclass[amsmath,amssymb,aps,letterpaper,prl,showpacs,twocolumn]{revtex4}

\fi

\allowdisplaybreaks[3]

\usepackage{fixmath}

\RequirePackage{verbatim}

\makeatletter
\newwrite\bibinl@out
\newenvironment{bibtex}[1][\jobname]{%
  \immediate\openout\bibinl@out #1.bib
  \immediate\write\bibinl@out{\@percentchar generated from `\jobname' starting line \the\inputlineno^^J}%
  \def\verbatim@processline{\immediate\write\bibinl@out{\the\verbatim@line}}%
  \@bsphack\let\do\@makeother\dospecials\catcode`\^^M\active\verbatim@start
}%
{\immediate\closeout\bibinl@out\@esphack}
\makeatother

\newcommand{\ham}{\mathcal{H}}
\newcommand{\charge}{\mathcal{Q}}
\newcommand{\yang}{\mathcal{Y}}
\newcommand{\loc}{\mathcal{L}}
\newcommand{\boost}[1]{\mathcal{B}[#1]}
\newcommand{\biloc}[2]{[#1|#2]}
\newcommand{\op}[1]{\mathcal{#1}}
\newcommand{\superN}{\mathcal{N}}

\newcommand{\order}[1]{\mathcal{O}(#1)}

\ifx\genfrac\sdflkaj

\else

\fi
\newcommand{\sfrac}[2]{{\textstyle\frac{#1}{#2}}}
\newcommand{\half}{\sfrac{1}{2}}
\newcommand{\ihalf}{\sfrac{i}{2}}

\newcommand{\supup}[1]{^{\mathrm{#1}}}

\newcommand{\lrbrk}[1]{\left(#1\right)}
\newcommand{\bigbrk}[1]{\bigl(#1\bigr)}

\newcommand{\bigcomm}[2]{\big[#1,#2\big]}
\newcommand{\comm}[2]{[#1,#2]}

\newcommand{\acomm}[2]{\{#1,#2\}}

\newcommand{\PTerm}[1]{[#1]}

\newcommand{\state}[1]{\mathopen{|}#1\mathclose{\rangle}}

\newcommand{\alg}[1]{\mathfrak{#1}}
\newcommand{\grp}[1]{\mathrm{#1}}

\newcommand{\nn}{\nonumber}
\newcommand{\nln}{\nonumber\\}
\newcommand{\nl}[1][0pt]{\nonumber\\[#1]&\hspace{-4\arraycolsep}&\mathord{}}
\newcommand{\nlnum}{\\&\hspace{-4\arraycolsep}&\mathord{}}
\newcommand{\earel}[1]{\mathrel{}&\hspace{-2\arraycolsep}#1\hspace{-2\arraycolsep}&\mathrel{}}
\newcommand{\eq}{\earel{=}}

\def\[{\begin{equation}}
\def\]{\end{equation}}
\def\<{\begin{eqnarray}}
\def\>{\end{eqnarray}}

\makeatletter
\def\mr@ignsp#1 {\ifx\:#1\@empty\else #1\expandafter\mr@ignsp\fi}%
\newcommand{\multiref}[1]{\begingroup
\xdef\mr@no@sparg{\expandafter\mr@ignsp#1 \: }%
\def\mr@comma{}%
\@for\mr@refs:=\mr@no@sparg\do{\mr@comma\def\mr@comma{,}\ref{\mr@refs}}%
\endgroup}
\makeatother

\renewcommand{\eqref}[1]{(\multiref{#1})}

\providecommand{\href}[2]{#2}
\newcommand{\arxivlink}[1]{\href{http://arxiv.org/abs/#1}{arxiv:#1}}

\begin{document}

\ifarxiv
\pdfbookmark[1]{Title Page}{title}
\begin{flushright}\footnotesize
\texttt{\arxivlink{0807.5081}}\\
\texttt{AEI-2008-052}%
\end{flushright}
\vspace{1cm}

\begin{center}%
{\Large\textbf{\mathversion{bold}%
Boosting Nearest-Neighbour to Long-Range Integrable Spin Chains}\par}
\vspace{1cm}%

\textsc{Till Bargheer, Niklas Beisert, Florian Loebbert}\vspace{5mm}%

\textit{Max-Planck-Institut f\"ur Gravitationsphysik\\%
Albert-Einstein-Institut\\%
Am M\"uhlenberg 1, 14476 Potsdam, Germany}\vspace{3mm}%

\texttt{till,nbeisert,florian.loebbert@aei.mpg.de}

\par\vspace{1cm}

\textbf{Abstract}\vspace{7mm}

\begin{minipage}{12.7cm}
We present an integrability-preserving recursion relation for the explicit construction  
of long-range spin chain Hamiltonians.
These chains are generalisations of the Haldane--Shastry and Inozemtsev 
models and they play an important role in recent advances in string/gauge duality.
The method is based on arbitrary nearest-neighbour integrable spin chains
and it sheds light on the moduli space of deformation parameters.
We also derive the closed chain asymptotic Bethe equations.
\end{minipage}

\end{center}


\vspace{1cm}
\hrule height 0.75pt
\vspace{1cm}

\fi

\section{Introduction and Overview}

Integrable spin chains are a fascinating topic of theoretical physics: 
Their Hilbert space grows exponentially with the length of the chain
and the Hamiltonian eigenstates are usually
in a linear combination of almost all states in a canonical basis. 
Nevertheless, owing to integrability, the eigenstates can be determined efficiently
by solving a system of algebraic equations,
the so-called Bethe equations \cite{Bethe:1931hc},
whose number of unknowns typically grows linearly with the length.

The best-studied integrable spin chains
are the \emph{nearest-neighbour} chains
whose Hamiltonians act on pairs of spins at adjacent sites.
The prime example in this class is
the Heisenberg model \cite{Heisenberg:1928aa}. 
The only widely-known examples of spin chains 
with interactions of well-separated spin sites, 
so-called \emph{long-range} chains,
are the Haldane--Shastry 
and the Inozemtsev chains \cite{Haldane:1988gg,SriramShastry:1988gh,Inozemtsev:1989yq,Inozemtsev:2002vb}. 

The discovery and investigation of integrable structures in planar
maximally supersymmetric gauge theory 
in four spacetime dimensions \cite{Minahan:2002ve,Beisert:2003tq,Beisert:2003yb}
(see \cite{Beisert:2004ry} for a review)
introduced a novel exciting long-range chain.
Its interactions are more general than those
on which the Haldane--Shastry and Inozemtsev models are based:
they involve more than two spins at a time.
A subsequent study \cite{Beisert:2005wv} has revealed 
a large class of long-range models.
However, in all of these models complete integrability
was merely shown to be plausible but has not yet been proven.

In this letter we provide a proof of existence
for a very large class of \emph{integrable} long-range spin chains
including those proposed in \cite{Beisert:2005wv}.
Specifically, we present a recursion relation 
to explicitly construct long-range Hamiltonians 
which manifestly preserves integrability.

\section{Integrable Long-Range Chains}

We start by reviewing the notion of perturbatively long-range integrable 
spin chain models \cite{Beisert:2003tq}.

\paragraph{General Definition.}

A perturbatively long-range integrable spin chain is defined as a deformation
of an infinitely long homogeneous nearest-neighbour integrable spin chain.
This means that the model has a set of local homogeneous 
commuting charges $\charge_r(\lambda)$, $r\geq 2$,
\[\label{eq:commuting}
\bigcomm{\charge_r(\lambda)}{\charge_s(\lambda)}=0,
\]
which are deformations of the commuting charges $\charge_r\supup{NN}$
of the nearest-neighbour model at $\lambda=0$.
For simplicity we shall identify the Hamiltonian 
$\ham$ with the lowest charge $\charge_2$.
The charges are composed from local operators
\[\label{eq:charges}
\charge_r(\lambda)=\sum\nolimits_k c_{r,k}(\lambda)\,\loc_k,
\]
where the $\loc_k$ form a basis of operators acting
\emph{locally} and \emph{homogeneously} on the chain
\[\label{eq:defloc}
\loc_k:=\sum\nolimits_{a}\loc_k(a).
\]
Here and in the following $\loc_k(a)$ is some operator which acts on
several consecutive spin sites starting with site $a$.
The number of interacting sites is called 
the range $[\loc_k]$ of the operator $\loc_k$.
The coefficients $c_{r,k}(\lambda)$ are defined 
as series expansions around $\lambda=0$ such that
the range of $\charge_r$ grows at most by one step per
order in $\lambda$, i.e.
\[\label{eq:range}
c_{r,k}(\lambda)=\order{\lambda^{[\loc_k]-r}}.
\]
In other words, $\charge_r(\lambda)$ at order $\lambda^\ell$ 
must consist of operators $\loc_k$ 
of range $[\loc_k]$ at most $r+\ell$.

\paragraph{Fundamental $\alg{gl}(N)$ Chain.}

The existence of interesting non-trivial integrable long-range models 
was suggested by the construction in \cite{Beisert:2003tq}. 
In fact, the hyperbolic Inozemtsev chain \cite{Inozemtsev:1989yq,Inozemtsev:2002vb} 
can be understood as one particular example \cite{Serban:2004jf}.
However, a general survey \cite{Beisert:2005wv}
of long-range chains with $\alg{gl}(N)$ symmetry 
and spins transforming in the fundamental representation 
has revealed a much larger moduli space:
The starting point was the ansatz \eqref{eq:charges,eq:range}
for the charges $\charge_{r}(\lambda)$, $r=2,3$, up to order $\order{\lambda^4}$.
The coefficients $c_{r,k}(\lambda)$ were then constrained by 
demanding commutativity \eqref{eq:commuting}.
The resulting first few charges at the leading few orders read
\<\label{eq:chargesGLN}
\charge_2\eq
\PTerm{1}-\PTerm{2,1}
\nl
+\alpha_3(\lambda) \bigbrk{-3\PTerm{1}+4\PTerm{2,1}-\PTerm{3,2,1}}
+\order{\lambda^2},
\nln
\charge_3\eq
\ihalf\bigbrk{\PTerm{3,1,2}-\PTerm{2,3,1}}
\nl
+\ihalf\alpha_3(\lambda) 
\bigl(6\PTerm{2,3,1}-6\PTerm{3,1,2}
+\PTerm{4,1,3,2}
\nl\,\,
+\PTerm{4,2,1,3}
-\PTerm{2,4,3,1}
-\PTerm{3,2,4,1}
\bigr)
+\order{\lambda^2},
\nln
\charge_4\eq
\sfrac{1}{3}
\bigl(
-\PTerm{1}
+2\PTerm{2,1}
-\PTerm{3,2,1}
+\PTerm{2,3,4,1}
\nlnum\nn\,\,
-\PTerm{2,4,1,3}
-\PTerm{3,1,4,2}
+\PTerm{4,1,2,3}
\bigr)
+\order{\lambda}.
\>
The symbols $\loc_k=\PTerm{\ldots}$ represent local homogeneous interactions
such that $\loc_k(a)$ in \eqref{eq:defloc} is the indicated permutation of 
consecutive spins, cf.\ \cite{Beisert:2007jv}. 
For example, $\PTerm{2,1}$ represents
the nearest-neighbour permutation $\sum_{a} P_{a,a+1}$.
The commuting charges turned out to depend on a set of parameters $\alpha_r$, 
$\beta_{r,s}$, $\gamma_{r,s}$ and $\epsilon_k$ whose
individual roles can be identified in the resulting Bethe ansatz.
The asymptotic Bethe equations for this model \cite{Beisert:2005wv} 
are a special case of the form presented at the end of this letter. 

Clearly this construction is sufficient 
neither to prove integrability at a certain perturbative order
nor to show that the deformation can be continued 
to higher orders without having to spoil integrability.
The first problem was overcome in \cite{Beisert:2007jv} 
by showing that $\alg{gl}(N)$ symmetry extends to a Yangian algebra. 
A perturbative Yangian generator $\yang(\lambda)$ was constructed,
shown to commute with $\charge_2(\lambda)$ and to satisfy the Serre
relations of the Yangian. 

It is the aim of the present letter to overcome both problems, namely
to show that the long-range integrable model can be constructed to all orders 
(and how).

\section{General Construction}

In the following we shall present the construction 
of long-range integrable spin chains 
from an arbitrary conventional integrable spin chain.

\paragraph{Generating Equation.}

Consider a one-parameter family 
of charges $\charge_r(\lambda)$
which obeys the differential equation
\[\label{eq:eq}
\frac{d}{d\lambda}\,\charge_r(\lambda)
=i\bigcomm{\op{X}(\lambda)}{\charge_r(\lambda)}.
\]
Here $\op{X}$ is some operator with well-defined
commutation relations with $\charge_r$ at all $\lambda$.
The differential equation guarantees that the algebra 
of the $\charge_r$ is independent of $\lambda$
%
\[
\frac{d}{d\lambda}\,\bigcomm{\charge_r(\lambda)}{\charge_s(\lambda)}
=i\bigcomm{\op{X}(\lambda)}{\comm{\charge_r(\lambda)}{\charge_s(\lambda)}}.
\]
In particular, if the algebra of charges is abelian anywhere,
e.g.\ at $\lambda=0$,
it is abelian everywhere
\[
\bigcomm{\charge_r(0)}{\charge_s(0)}=0 
\quad\Longrightarrow\quad
\bigcomm{\charge_r(\lambda)}{\charge_s(\lambda)}=0. 
\]
Given a conventional nearest-neighbour integrable system 
$\charge\supup{NN}_r$ and some operator $\op{X}(\lambda)$, 
the differential equation \eqref{eq:eq}
defines an integrable deformation 
$\charge_r(\lambda)$ of $\charge_r(0)=\charge\supup{NN}_r$, 
at least as a formal series in $\lambda$.

The integrable charges $\charge_r(\lambda)$ 
of the long-range model 
discussed above are \emph{local}
and \emph{homogeneous}.
Consequently, if \eqref{eq:eq} is to describe the above model
we have to make sure that the equation violates neither of these properties.
More explicitly, $\comm{\op{X}}{\charge_r}$ must be local and homogeneous
for all $\lambda$. In the following we shall discuss
suitable choices for $\op{X}$.

\paragraph{Local Operators.}

Obviously, the commutator of any two local operators
is again local. Thus any local operator is admissible
as a deformation and we can set
\[\label{eq:localdef}
\op{X}(\lambda)=\sum\nolimits_k \epsilon_k(\lambda)\,\loc_k+\ldots,
\]
where the $\loc_k$ form a basis of local operators.
This deformation changes eigenstates only locally 
and has no impact on the spectrum.
The $\epsilon$'s are thus unphysical.
Note that for a correct enumeration of deformation degrees of freedom
one has to take into account that the 
local charges $\charge_r(\lambda)$ generate
trivial deformations.

\paragraph{Boost Charges.}

Consider \emph{boost operators} defined by 
\[\label{eq:defboost}
\boost{\loc_k}:=\sum\nolimits_{a} a\,\loc_k(a).
\]
In contradistinction to the $\loc_k$ defined in 
\eqref{eq:defloc} a boost acts locally,
but inhomogeneously along the chain.
Recall \cite{Tetelman:1981xx,Sogo:1983aa}
that the boost $\boost{\charge_2}$
can be used to generate all the higher charges of a
conventional integrable system through the recursive relation
$i\bigcomm{\boost{\charge_2}}{\charge_r}\simeq -r\charge_{r+1}$.

In general, the commutator of some boost operator 
with some local operator is again a boost operator,
$\comm{\boost{\loc_k}}{\loc_l}=\boost{\loc_m}$.
However, if the underlying local operators commute,
the commutator becomes homogeneous
\[
\comm{\loc_k}{\loc_l}=0\quad
\Longrightarrow\quad
\bigcomm{\boost{\loc_k}}{\loc_l}=\loc_m.
\]
Consequently the boosts of commuting charges are admissible
as deformations
\[\label{eq:boostdef}
\op{X}(\lambda)=\ldots+\sum_{r=3}^\infty \tilde\alpha_r(\lambda)\,\boost{\charge_r(\lambda)}+\ldots. 
\]
Note that the definition of $\boost{\loc_k}$
is ambiguous modulo local operators $\loc_l$.
This is not troublesome because we 
have already accounted for all local operators in \eqref{eq:localdef},
i.e.\ the unphysical parameters $\epsilon_k$ can absorb the ambiguity.


\paragraph{Bi-Local Charges.}

Next, consider bi-local operators 
\[\label{eq:defbiloc}
\biloc{\loc_k}{\loc_l}:=
\sum\nolimits_{a\leq b} \half(1-\half\delta_{a,b})\acomm{\loc_k(a)}{\loc_l(b)}.
\]
The commutation properties of bi-local operators
$\biloc{\loc_k}{\loc_l}$
are reminiscent of those of boost operators discussed above.
A commutator of a bi-local with a local operator yields
a bi-local operator in general. However, for commuting charges
it remains local 
\[
\comm{\loc_{k,l}}{\loc_m}=0\quad
\Longrightarrow\quad
\bigcomm{\biloc{\loc_k}{\loc_l}}{\loc_m}=\loc_n.
\]
Therefore bi-local combinations of the charges are admissible
as deformations
\[
\op{X}(\lambda)=\ldots+\sum_{s>r=2}^\infty \tilde\beta_{r,s}(\lambda)\,
\biloc{\charge_r(\lambda)}{\charge_s(\lambda)}.
\]
Note that also bi-local operators are uniquely defined only modulo local operators
which is alright due to \eqref{eq:localdef}.

\paragraph{Shifts.}

The above operators exhaust all admissible operators we can think of. 
In fact they almost agree with the proposed moduli space
of the Bethe ansatz \cite{Beisert:2005wv}. 
The only missing parameters
correspond to taking linear combinations of the charges. 
We introduce them by adding another
term to the equation \eqref{eq:eq} 
which obviously does not spoil integrability
\[\label{eq:eq2}
\frac{d}{d\lambda}\,\charge_r(\lambda)
=i\bigcomm{\op{X}(\lambda)}{\charge_r(\lambda)}
+\sum_{s=2}^\infty \tilde\gamma_{r,s}(\lambda)\,\charge_s(\lambda).
\]
%

\paragraph{Yangian Generators.}

The Yangian generators are deformed 
in the same way as the integrable charges
\[\label{eq:eqYang}
\frac{d}{d\lambda}\,\yang(\lambda)
=i\bigcomm{\op{X}(\lambda)}{\yang(\lambda)}.
\]
This guarantees that the algebra among the Yangian 
generators and the integrable charges is 
the same for all values of $\lambda$.
In particular it is the same as for 
the conventional integrable chain at $\lambda=0$
in line with the results \cite{Beisert:2007jv} 
on the Yangian of $\alg{gl}(N)$ long-range chains.

\section{Interaction Range}

The above definition of long-range chains \eqref{eq:range}
sets the bound $r+\ell$ to the range of interactions 
in $\charge_r$ at $\order{\lambda^\ell}$.
A superficial consideration of the range of 
commutators in \eqref{eq:eq} shows that 
each power $\tilde\alpha_r$ 
increases the range of a charge $\charge$ by $r-1$. 
Likewise, $\tilde\beta_{r,s}$ naively increases the range by $r+s-2$.
However, we observe \cite{Beisert:2005wv}, cf.\ \eqref{eq:chargesGLN},
that $\alpha_r$ merely increases the range by $r-2$
and $\beta_{r,s}$ by $s-1$.

\paragraph{Boost Charges.}

Let us consider boost charges first, which superficially generate
terms too long by one site.
We observe that the contributions of leading range 
agree precisely with those in some conserved charge
\[
i(s-1)\bigcomm{\boost{\charge_s(\lambda)}}{\charge_r(\lambda)} \simeq 
-(s+r-2)\charge_{s+r-1}.
\]
Consequently we can reduce the range by one site
by fixing $\tilde\gamma_{r,s}$ appropriately
($\tilde\gamma_{r,s}=0$ for $s<r+2$)
\[\label{eq:defgamma}
\tilde\gamma_{r,s}(\lambda)=\frac{s-1}{s-r}\,\tilde\alpha_{s-r+1}(\lambda).
\]

Furthermore, combinations of multiple $\tilde\alpha$'s 
lead to a range which is longer than expected. 
This problem can apparently be cured by choosing 
the $\tilde\alpha$'s as follows
\[\label{eq:atofu}
\sum_{r=3}^\infty \frac{\tilde\alpha_r(\lambda)}{(r-1)x^{r-2}}
=\frac{du(x)}{d\lambda}\bigg/\frac{du(x)}{dx}
=-\frac{dx(u)}{d\lambda}\,.
\]
where $u(x)$ is a function of refined $\alpha$'s 
\[\label{eq:uofx}
u(x)=x+\sum_{r=3}^\infty\frac{\alpha_r(\lambda)}{x^{r-2}}\,,
\qquad 
\alpha_r(\lambda)=\order{\lambda^{r-2}},
\]
and $x(u)=u+\order{\lambda}$ is its inverse.
Presently, we have no good understanding 
of why the subtraction \eqref{eq:defgamma}
and the function \eqref{eq:atofu,eq:uofx} reduce
the range or how to prove these observations.

\paragraph{Bi-Local Charges.}

Similarly, the range of the terms due to bi-local charges
depends on the definition of $\tilde\beta$'s.
The correct choice seems to be (with $\beta_{r,s}=-\beta_{s,r}$)
\<
\tilde\beta_{r,s}(\lambda)\eq
2\beta'_{r,s}(\lambda)
+\sum_{r'=2}^{r-2} 2\,\beta_{r',s}(\lambda)\,\tilde\gamma_{r',r}(\lambda)
\nln\earel{}\phantom{2\beta'_{r,s}(\lambda)}
+\sum_{s'=2}^{s-2} 2\,\beta_{r,s'}(\lambda)\,\tilde\gamma_{s',s}(\lambda).
\>
Furthermore the regularisation of bi-local operators
in \eqref{eq:defbiloc}
apparently reduces the range as far as possible, i.e.
\[
\beta_{r,s}(\lambda)=\order{\lambda^{s-1}}.
\]
%

\section{Bethe Ansatz}

We would now like to apply the coordinate Bethe ansatz 
in order to derive the asymptotic Bethe equations.

\paragraph{Dispersion Relations.}

First we prepare a one-magnon eigenstate $\state{p}$ with definite momentum $p$
along the chain in order to measure the dispersion relations
of the charges. Applying the recursion relation 
\eqref{eq:eq2} to the state yields an equation 
on the charge eigenvalues $q_r$
\[
\frac{dq_r}{d\lambda}
=-\sum_{s=3}^\infty\tilde\alpha_s q_s \frac{dq_r}{dp}+
\sum_{s=2}^\infty \tilde\gamma_{r,s} q_{s}.
\]
A solution with integration constant $t$ reads,
cf.\ \cite{Beisert:2004hm}
\[\label{eq:defq}
q_{r}(t,u)=\frac{i}{r-1}\lrbrk{\frac{1}{x(u+\ihalf t)^{r-1}}-\frac{1}{x(u-\ihalf t)^{r-1}}}.
\]
The rapidity $u(p)$ is defined implicitly by
\[
\exp\bigbrk{ip(t,u)}=\frac{x(u+\ihalf t)}{x(u-\ihalf t)}\,.
\]

\paragraph{Dressing Phase.}

To understand how the S-matrix $S$
depends on $\lambda$ we consider two-magnon eigenstates
\[
\state{u,u'} = \state{u<u'}+S(u,u')\state{u'<u}+\state{\mathrm{local}}.
\]
The magnon momenta $p,p'$ are implicitly defined 
through the rapidities $u,u'$.
The recursion relation \eqref{eq:eq2} implies the following dependence
\[
S(u,u')=\exp\bigbrk{-2i\theta(u,u')}S_0(u,u'),
\]
where $S_0(u,u')$ is the scattering matrix at $\lambda=0$.
The dressing phase takes the form proposed in \cite{Arutyunov:2004vx,Beisert:2005wv}
\[
\theta(u,u')=
\sum_{s>r=2}^\infty\beta_{r,s}\bigbrk{q_r(u)\,q_s(u') - q_r(u')\,q_s(u)}.
\]
%

\paragraph{Bethe Equations.}

Based on the above results for the dispersion relations and
the scattering matrix we can write the deformed 
Bethe equations. 
The Bethe equations for an integrable spin chain 
based on an R-matrix with Yangian symmetry $\grp{Y}(\alg{g})$
have been developed in \cite{Reshetikhin:1983vw,Reshetikhin:1985vd,Ogievetsky:1986hu}.
The Lie (super)algebra $\alg{g}$ of rank $R$ is
specified by the symmetric Cartan matrix $C_{ab}$, $a,b=1,\ldots,R$.
The closed spin chain consists of $L$ identical Yangian modules
with Dynkin labels $t_a$, $a=1,\ldots,R$.
Periodic eigenstates of the spin chain
are described by the Bethe roots $u_{a,k}$, $k=1,\ldots,K_a$,
satisfying the Bethe equations
\<
\earel{}\exp\bigbrk{ip(t_a,u_{a,k})L}
\\\nn
\eq
\hspace{-0.2cm}
\mathop{\prod_{b=1}^R\prod_{j=1}^{K_b}}_{(b,j)\neq(a,k)}
\hspace{-0.2cm}
\frac{u_{a,k}-u_{b,j}+\ihalf C_{ab}}{u_{a,k}-u_{b,j}-\ihalf C_{ab}}\,
\exp\bigbrk{2i\theta(t_a,u_{a,k};t_b,u_{b,j})}
.
\>
The charge eigenvalues then take the form
\<
e^{iP}\eq\prod_{a=1}^R\prod_{k=1}^{K_a}\exp\bigbrk{ip(t_a,u_{k,a})},
\nln
Q_r\eq\sum_{a=1}^R\sum_{k=1}^{K_a} q_{r}(t_a,u_{a,k}).
\>
Note that these Bethe equations are merely \emph{asymptotic}
\cite{Sutherland:1978aa}:
The charge eigenvalues $Q_r$ are valid only up to terms 
of order $\order{\lambda^{L-r+1}}$ for which the range of 
$\charge_r$ exceeds $L$ and where it is thus not properly defined.

\section{Conclusions and Outlook}

In this note we have presented a recursion relation to
construct integrable long-range spin chains from 
an arbitrary short-ranged model, cf.\ \cite{Bargheer:2009xy} for further details. 
These models have appeared in the context of
$\superN=4$ supersymmetric gauge theory, but their existence 
and all-orders consistency was largely conjectural so far.

The method applies to generic Lie (super)algebras 
and spin representations and it explains
the set of allowed deformation parameters.
The deformation parameters control the range of
interactions and by taking suitable combinations
of some parameters we were able to decrease the range systematically
and in accordance with planar gauge theory.
This observation, however, lacks a proof.
Our construction method applies to infinitely long chains,
but using the coordinate Bethe ansatz and earlier 
quantum algebra results we have derived 
asymptotic Bethe equations for closed chains.

Several aspects deserve further scrutiny:
It would be important to have a better
understanding of how and why the range is decreased
for boost deformations.
In particular, does the reduction apply to all
algebras, representations and to quantum deformations?
For example, we find \cite{Bargheer:2009xy}  
that it apparently applies to alternating spin chains 
as the ones recently found \cite{Minahan:2008hf} (see also \cite{Bak:2008cp}) 
in $\superN=6$ superconformal Chern--Simons theory \cite{Aharony:2008ug}.

The moduli space of open long-range integrable chains 
is slightly different \cite{Beisert:2008cf}, e.g.\ a phase
associated to the boundaries appears. 
Can such open chains including new degrees
of freedom be constructed in a similar fashion?

In our model the Lie algebra symmetry is manifest
whereas for $\superN=4$ gauge theory
only the compact part of $\alg{psu}(2,2|4)$ acts canonically.
The other generators
are deformed much alike the Hamiltonian,
which is an inseparable part of the algebra.
Recently a very similar recursion relation to ours has 
appeared for such systems \cite{Zwiebel:2008gr} 
and it would be highly desirable to join the two structures
to construct the complete $\alg{psu}(2,2|4)$ representation.


\begin{bibtex}
@Article{Aharony:2008ug,
     author    = "Aharony, Ofer and Bergman, Oren and Jafferis, Daniel Louis
                  and Maldacena, Juan",
     title     = "$\mathcal{N}$ = 6 superconformal Chern-Simons-matter theories, M2-branes
                  and their gravity duals",
      journal        = "JHEP",
      volume         = "0810",
      pages          = "091",
      doi            = "10.1088/1126-6708/2008/10/091",
      year           = "2008",
     eprint    = "0806.1218",
     archivePrefix = "arXiv",
     primaryClass  =  "hep-th",
     SLACcitation  = "
}

@Article{Arutyunov:2004vx,
     author    = "Arutyunov, Gleb and Frolov, Sergey and Staudacher, Matthias",
     title     = "Bethe ansatz for quantum strings",
     journal   = "JHEP",
     volume    = "0410",
     year      = "2004",
     pages     = "016",
     eprint    = "hep-th/0406256",
     doi       = "10.1088/1126-6708/2004/10/016",
     SLACcitation  = "
}

@Article{Bak:2008cp,
     author    = "Bak, Dongsu and Rey, Soo-Jong",
     title     = "Integrable Spin Chain in Superconformal Chern-Simons
                  Theory",
      journal        = "JHEP",
      volume         = "0810",
      pages          = "053",
      doi            = "10.1088/1126-6708/2008/10/053",
      year           = "2008",
     eprint    = "0807.2063",
     archivePrefix = "arXiv",
     primaryClass  =  "hep-th",
     SLACcitation  = "
}

@article{Bargheer:2009xy,
      author         = "Bargheer, Till and Beisert, Niklas and Loebbert, Florian",
      title          = "Long-Range Deformations for Integrable Spin Chains",
      journal        = "J.Phys.",
      volume         = "A42",
      pages          = "285205",
      doi            = "10.1088/1751-8113/42/28/285205",
      year           = "2009",
      eprint         = "0902.0956",
      archivePrefix  = "arXiv",
      primaryClass   = "hep-th",
      reportNumber   = "AEI-2009-009",
      SLACcitation   = "
}

@Article{Beisert:2003tq,
     author    = "Beisert, N. and Kristjansen, C. and Staudacher, M.",
     title     = "The Dilatation Operator of $\mathcal{N}$ = 4 Conformal Super Yang-Mills Theory",
     journal   = "Nucl. Phys.",
     volume    = "B664",
     year      = "2003",
     pages     = "131-184",
     eprint    = "hep-th/0303060",
     doi       = "10.1016/S0550-3213(03)00406-1",
     SLACcitation  = "
}

@Article{Beisert:2003yb,
     author    = "Beisert, Niklas and Staudacher, Matthias",
     title     = "The $\mathcal{N}$ = 4 SYM Integrable Super Spin Chain",
     journal   = "Nucl. Phys.",
     volume    = "B670",
     year      = "2003",
     pages     = "439-463",
     eprint    = "hep-th/0307042",
     doi       = "10.1016/j.nuclphysb.2003.08.015",
     SLACcitation  = "
}

@Article{Beisert:2004hm,
     author    = "Beisert, N. and Dippel, V. and Staudacher, M.",
     title     = "A Novel Long Range Spin Chain and Planar $\mathcal{N}$ = 4 Super Yang-Mills",
     journal   = "JHEP",
     volume    = "0407",
     year      = "2004",
     pages     = "075",
     eprint    = "hep-th/0405001",
     doi       = "10.1088/1126-6708/2004/07/075",
     SLACcitation  = "
}

@Article{Beisert:2004ry,
     author    = "Beisert, Niklas",
     title     = "The Dilatation Operator of $\mathcal{N}$ = 4 Super Yang-Mills Theory and Integrability",
     journal   = "Phys. Rept.",
     volume    = "405",
     year      = "2004",
     pages     = "1-202",
     eprint    = "hep-th/0407277",
     doi       = "10.1016/j.physrep.2004.09.007",
     SLACcitation  = "
}

@Article{Beisert:2005wv,
     author    = "Beisert, N. and Klose, T.",
     title     = "Long-Range GL(n) Integrable Spin Chains and Plane-Wave Matrix Theory",
     journal   = "J. Stat. Mech.",
     volume    = "06",
     year      = "2006",
     pages     = "P07006",
     doi       = "10.1088/1742-5468/2006/07/P07006",
     eprint    = "hep-th/0510124",
     SLACcitation  = "
}

@Article{Beisert:2007jv,
     author    = "Beisert, Niklas and Erkal, Denis",
     title     = "Yangian Symmetry of Long-Range gl(N) Integrable Spin
                  Chains",
     journal   = "J. Stat. Mech.",
     volume    = "08",
     year      = "2008",
     pages     = "P03001",
     eprint    = "0711.4813",
     doi       = "10.1088/1742-5468/2008/03/P03001",
     SLACcitation  = "
}

@Article{Beisert:2008cf,
     author    = "Beisert, N. and Loebbert, F.",
     title     = "Open Perturbatively Long-Range Integrable gl(N) Spin
                  Chains",
      journal        = "Adv.Sci.Lett.",
      volume         = "2",
      pages          = "261-269",
      year           = "2009",
      eprint         = "0805.3260",
      doi = "10.1166/asl.2009.1034",
     archivePrefix = "arXiv",
     primaryClass  =  "hep-th",
     SLACcitation  = "
}

@Article{Bethe:1931hc,
     author    = "Bethe, H.",
     title     = "Zur Theorie der Metalle I. Eigenwerte und Eigenfunktionen der linearen Atomkette",
     journal   = "Z. Phys.",
     volume    = "71",
     year      = "1931",
     pages     = "205-226",
     doi       = "10.1007/BF01341708",
     SLACcitation  = "
}

@Article{Heisenberg:1928aa,
     author    = "Heisenberg, W.",
     title     = "Zur Theorie des Ferromagnetismus",
     journal   = "Z. Phys.",
     volume    = "49",
     year      = "1928",
     pages     = "619-636",
     doi       = "10.1007/BF01328601",
     SLACcitation  = "
}

@Article{Haldane:1988gg,
     author    = "Haldane, F. D. M.",
     title     = "Exact Jastrow-Gutzwiller resonating valence bond ground
                  state of the spin $1/2$ antiferromagnetic Heisenberg chain
                  with $1/r^2$ exchange",
     journal   = "Phys. Rev. Lett.",
     volume    = "60",
     year      = "1988",
     pages     = "635",
     doi       = "10.1103/PhysRevLett.60.635",
     SLACcitation  = "
}

@Article{Inozemtsev:1989yq,
     author    = "Inozemtsev, V. I.",
     title     = "On the connection between the one-dimensional s = $1/2$
                  Heisenberg chain and Haldane Shastry model",
     journal   = "J. Stat. Phys.",
     volume    = "59",
     year      = "1990",
     pages     = "1143",
     doi = "10.1007/BF01334745",
}

@Article{Inozemtsev:2002vb,
     author    = "Inozemtsev, V. I.",
     title     = "Integrable Heisenberg-van Vleck chains with variable range
                  exchange",
     journal   = "Phys. Part. Nucl.",
     volume    = "34",
     year      = "2003",
     pages     = "166-193",
     eprint    = "hep-th/0201001",
     SLACcitation  = "
}

@Article{Minahan:2002ve,
     author    = "Minahan, J. A. and Zarembo, K.",
     title     = "The Bethe-ansatz for $\mathcal{N}$ = 4 super Yang-Mills",
     journal   = "JHEP",
     volume    = "0303",
     year      = "2003",
     pages     = "013",
     eprint    = "hep-th/0212208",
     doi       = "10.1088/1126-6708/2003/03/013",
     SLACcitation  = "
}

@Article{Minahan:2008hf,
     author    = "Minahan, J. A. and Zarembo, K.",
     title     = "{The Bethe ansatz for superconformal Chern-Simons}",
      journal        = "JHEP",
      volume         = "0809",
      pages          = "040",
      doi            = "10.1088/1126-6708/2008/09/040",
      year           = "2008",
     eprint    = "0806.3951",
     archivePrefix = "arXiv",
     primaryClass  =  "hep-th",
     SLACcitation  = "
}

@Article{Ogievetsky:1986hu,
     author    = "Ogievetsky, E. and Wiegmann, P.",
     title     = "Factorized S matrix and the Bethe ansatz for simple Lie groups",
     journal   = "Phys. Lett.",
     volume    = "B168",
     year      = "1986",
     pages     = "360",
     doi       = "10.1016/0370-2693(86)91644-8",
     SLACcitation  = "
}

@Article{Reshetikhin:1983vw,
     author    = "Reshetikhin, N. {\relax Yu}.",
     title     = "A method of functional equations in the theory
                  of exactly solvable quantum system",
     journal   = "Lett. Math. Phys.",
     volume    = "7",
     year      = "1983",
     pages     = "205-213",
     doi       = "10.1007/BF00400435",
     SLACcitation  = "
}

@Article{Reshetikhin:1985vd,
     author    = "Reshetikhin, N. {\relax Yu}.",
     title     = "Integrable models of quantum one-dimensional magnets
                  with O(N) and Sp(2K) symmetry",
     journal   = "Theor. Math. Phys.",
     volume    = "63",
     year      = "1985",
     pages     = "555-569",
     doi       = "10.1007/BF01017501",
     SLACcitation  = "
}

@Article{Serban:2004jf,
     author    = "Serban, Didina and Staudacher, Matthias",
     title     = "Planar $\mathcal{N}$ = 4 gauge theory
                  and the Inozemtsev long range spin chain",
     journal   = "JHEP",
     volume    = "0406",
     year      = "2004",
     pages     = "001",
     eprint    = "hep-th/0401057",
     doi       = "10.1088/1126-6708/2004/06/001",
     SLACcitation  = "
}

@Article{Sogo:1983aa,
   author = "Sogo, K. and Wadati, M.",
    title = "Boost Operator and Its Application to Quantum Gelfand-Levitan Equation for Heisenberg-Ising Chain with Spin One-Half",
  journal = "Prog. Theor. Phys.",
     year = "1983",
   volume = "69",
    pages = "431-450",
    doi   = "10.1143/PTP.69.431"
}

@Article{SriramShastry:1988gh,
     author    = "Shastry, B. S.",
     title     = "Exact solution of an S = $1/2$ Heisenberg antiferromagnetic
                  chain with long ranged interactions",
     journal   = "Phys. Rev. Lett.",
     volume    = "60",
     year      = "1988",
     pages     = "639",
     doi       = "10.1103/PhysRevLett.60.639",
     SLACcitation  = "
}

@Article{Sutherland:1978aa,
     author    = "Sutherland, Bill",
     title     = "A brief history of the quantum soliton with new results on
                 the quantization of the Toda lattice",
     journal   = "Rocky Mtn. J. Math.",
     volume    = "8",
     year      = "1978",
     pages     = "413"
}

@Article{Zwiebel:2008gr,
     author    = "Zwiebel, Benjamin I.",
     title     = "Iterative Structure of the $\mathcal{N}$ = 4 SYM Spin Chain",
      journal        = "JHEP",
      volume         = "0807",
      pages          = "114",
      doi            = "10.1088/1126-6708/2008/07/114",
      year           = "2008",
     eprint    = "0806.1786",
     archivePrefix = "arXiv",
     primaryClass  =  "hep-th",
     SLACcitation  = "
}

@Article{Tetelman:1981xx,
      author    = "M. G. Tetel'man",
      title     = "{Lorentz group for two-dimensional integrable lattice systems}",
      journal   = "Sov. Phys. JETP.",
      volume    = "55",
      year      = "1982",
      pages     = "306",
      SLACcitation  = "
}
\end{bibtex}

\pdfbookmark[1]{\refname}{references}
\bibliography{boost}

\begin{thebibliography}{10}
\providecommand{\href}[2]{#2}
\providecommand{\arxivref}[2]{\href{http://arxiv.org/abs/#1}{#2}}
\providecommand{\doiref}[2]{\href{http://dx.doi.org/#1}{#2}}
\providecommand{\nbbstauthor}[1]{#1}
\providecommand{\nbbstjournal}[1]{\textsf{#1}}
\providecommand{\nbbsttitle}[1]{\textit{#1}}
\providecommand{\nbbsturl}[1]{\texttt{#1}}
\providecommand{\nbbsteprint}[1]{\texttt{#1}}
\providecommand{\nbbststyle}{\raggedright\small\parskip0pt}
\nbbststyle

\bibitem{Bethe:1931hc}
\nbbstauthor{H.~Bethe},
\nbbsttitle{``Zur Theorie der Metalle I. Eigenwerte und Eigenfunktionen der
  linearen Atomkette''},
\nbbstjournal{\doiref{10.1007/BF01341708}{Z.~Phys.~71,~205~(1931)}}.

\bibitem{Heisenberg:1928aa}
\nbbstauthor{W.~Heisenberg},
\nbbsttitle{``Zur Theorie des Ferromagnetismus''},
\nbbstjournal{\doiref{10.1007/BF01328601}{Z.~Phys.~49,~619~(1928)}}.

\bibitem{Haldane:1988gg}
\nbbstauthor{F.~D.~M.~Haldane},
\nbbsttitle{``Exact Jastrow-Gutzwiller resonating valence bond ground state of
  the spin $1/2$ antiferromagnetic Heisenberg chain with $1/r^2$ exchange''},
\nbbstjournal{\doiref{10.1103/PhysRevLett.60.635}{Phys.~Rev.~Lett.~60,~635~(1988)}}.

\bibitem{SriramShastry:1988gh}
\nbbstauthor{B.~S.~Shastry},
\nbbsttitle{``Exact solution of an S = $1/2$ Heisenberg antiferromagnetic chain
  with long ranged interactions''},
\nbbstjournal{\doiref{10.1103/PhysRevLett.60.639}{Phys.~Rev.~Lett.~60,~639~(1988)}}.

\bibitem{Inozemtsev:1989yq}
\nbbstauthor{V.~I.~Inozemtsev},
\nbbsttitle{``On the connection between the one-dimensional s = $1/2$
  Heisenberg chain and Haldane Shastry model''},
\nbbstjournal{\doiref{10.1007/BF01334745}{J.~Stat.~Phys.~59,~1143~(1990)}}.

\bibitem{Inozemtsev:2002vb}
\nbbstauthor{V.~I.~Inozemtsev},
\nbbsttitle{``Integrable Heisenberg-van Vleck chains with variable range
  exchange''},
\nbbstjournal{Phys.~Part.~Nucl.~34,~166~(2003)},
\nbbsteprint{\arxivref{hep-th/0201001}{hep-th/0201001}}.

\bibitem{Minahan:2002ve}
\nbbstauthor{J.~A.~Minahan and K.~Zarembo},
\nbbsttitle{``The Bethe-ansatz for $\mathcal{N}$ = 4 super Yang-Mills''},
\nbbstjournal{\doiref{10.1088/1126-6708/2003/03/013}{JHEP~0303,~013~(2003)}},
\nbbsteprint{\arxivref{hep-th/0212208}{hep-th/0212208}}.

\bibitem{Beisert:2003tq}
\nbbstauthor{N.~Beisert, C.~Kristjansen and M.~Staudacher},
\nbbsttitle{``The Dilatation Operator of $\mathcal{N}$ = 4 Conformal Super
  Yang-Mills Theory''},
\nbbstjournal{\doiref{10.1016/S0550-3213(03)00406-1}{Nucl.~Phys.~B664,~131~(2003)}},
\nbbsteprint{\arxivref{hep-th/0303060}{hep-th/0303060}}.

\bibitem{Beisert:2003yb}
\nbbstauthor{N.~Beisert and M.~Staudacher},
\nbbsttitle{``The $\mathcal{N}$ = 4 SYM Integrable Super Spin Chain''},
\nbbstjournal{\doiref{10.1016/j.nuclphysb.2003.08.015}{Nucl.~Phys.~B670,~439~(2003)}},
\nbbsteprint{\arxivref{hep-th/0307042}{hep-th/0307042}}.

\bibitem{Beisert:2004ry}
\nbbstauthor{N.~Beisert},
\nbbsttitle{``The Dilatation Operator of $\mathcal{N}$ = 4 Super Yang-Mills
  Theory and Integrability''},
\nbbstjournal{\doiref{10.1016/j.physrep.2004.09.007}{Phys.~Rept.~405,~1~(2004)}},
\nbbsteprint{\arxivref{hep-th/0407277}{hep-th/0407277}}.

\bibitem{Beisert:2005wv}
\nbbstauthor{N.~Beisert and T.~Klose},
\nbbsttitle{``Long-Range GL(n) Integrable Spin Chains and Plane-Wave Matrix
  Theory''},
\nbbstjournal{\doiref{10.1088/1742-5468/2006/07/P07006}{J.~Stat.~Mech.~06,~P07006~(2006)}},
\nbbsteprint{\arxivref{hep-th/0510124}{hep-th/0510124}}.

\bibitem{Serban:2004jf}
\nbbstauthor{D.~Serban and M.~Staudacher},
\nbbsttitle{``Planar $\mathcal{N}$ = 4 gauge theory and the Inozemtsev long
  range spin chain''},
\nbbstjournal{\doiref{10.1088/1126-6708/2004/06/001}{JHEP~0406,~001~(2004)}},
\nbbsteprint{\arxivref{hep-th/0401057}{hep-th/0401057}}.

\bibitem{Beisert:2007jv}
\nbbstauthor{N.~Beisert and D.~Erkal},
\nbbsttitle{``Yangian Symmetry of Long-Range gl(N) Integrable Spin Chains''},
\nbbstjournal{\doiref{10.1088/1742-5468/2008/03/P03001}{J.~Stat.~Mech.~08,~P03001~(2008)}},
\nbbsteprint{\arxivref{0711.4813}{arxiv:0711.4813}}.

\bibitem{Tetelman:1981xx}
\nbbstauthor{M.~G.~Tetel'man},
\nbbsttitle{``{Lorentz group for two-dimensional integrable lattice
  systems}''},
\nbbstjournal{Sov.~Phys.~JETP.~55,~306~(1982)}.

\bibitem{Sogo:1983aa}
\nbbstauthor{K.~Sogo and M.~Wadati},
\nbbsttitle{``Boost Operator and Its Application to Quantum Gelfand-Levitan
  Equation for Heisenberg-Ising Chain with Spin One-Half''},
\nbbstjournal{\doiref{10.1143/PTP.69.431}{Prog.~Theor.~Phys.~69,~431~(1983)}}.

\bibitem{Beisert:2004hm}
\nbbstauthor{N.~Beisert, V.~Dippel and M.~Staudacher},
\nbbsttitle{``A Novel Long Range Spin Chain and Planar $\mathcal{N}$ = 4 Super
  Yang-Mills''},
\nbbstjournal{\doiref{10.1088/1126-6708/2004/07/075}{JHEP~0407,~075~(2004)}},
\nbbsteprint{\arxivref{hep-th/0405001}{hep-th/0405001}}.

\bibitem{Arutyunov:2004vx}
\nbbstauthor{G.~Arutyunov, S.~Frolov and M.~Staudacher},
\nbbsttitle{``Bethe ansatz for quantum strings''},
\nbbstjournal{\doiref{10.1088/1126-6708/2004/10/016}{JHEP~0410,~016~(2004)}},
\nbbsteprint{\arxivref{hep-th/0406256}{hep-th/0406256}}.

\bibitem{Reshetikhin:1983vw}
\nbbstauthor{N.~{\relax Yu}.~Reshetikhin},
\nbbsttitle{``A method of functional equations in the theory of exactly
  solvable quantum system''},
\nbbstjournal{\doiref{10.1007/BF00400435}{Lett.~Math.~Phys.~7,~205~(1983)}}.

\bibitem{Reshetikhin:1985vd}
\nbbstauthor{N.~{\relax Yu}.~Reshetikhin},
\nbbsttitle{``Integrable models of quantum one-dimensional magnets with O(N)
  and Sp(2K) symmetry''},
\nbbstjournal{\doiref{10.1007/BF01017501}{Theor.~Math.~Phys.~63,~555~(1985)}}.

\bibitem{Ogievetsky:1986hu}
\nbbstauthor{E.~Ogievetsky and P.~Wiegmann},
\nbbsttitle{``Factorized S matrix and the Bethe ansatz for simple Lie
  groups''},
\nbbstjournal{\doiref{10.1016/0370-2693(86)91644-8}{Phys.~Lett.~B168,~360~(1986)}}.

\bibitem{Sutherland:1978aa}
\nbbstauthor{B.~Sutherland},
\nbbsttitle{``A brief history of the quantum soliton with new results on the
  quantization of the Toda lattice''},
\nbbstjournal{Rocky~Mtn.~J.~Math.~8,~413~(1978)}.

\bibitem{Bargheer:2009xy}
\nbbstauthor{T.~Bargheer, N.~Beisert and F.~Loebbert},
\nbbsttitle{``Long-Range Deformations for Integrable Spin Chains''},
\nbbstjournal{\doiref{10.1088/1751-8113/42/28/285205}{J.~Phys.~A42,~285205~(2009)}},
\nbbsteprint{\arxivref{0902.0956}{arxiv:0902.0956}}.

\bibitem{Minahan:2008hf}
\nbbstauthor{J.~A.~Minahan and K.~Zarembo},
\nbbsttitle{``{The Bethe ansatz for superconformal Chern-Simons}''},
\nbbstjournal{\doiref{10.1088/1126-6708/2008/09/040}{JHEP~0809,~040~(2008)}},
\nbbsteprint{\arxivref{0806.3951}{arxiv:0806.3951}}.

\bibitem{Bak:2008cp}
\nbbstauthor{D.~Bak and S.-J.~Rey},
\nbbsttitle{``Integrable Spin Chain in Superconformal Chern-Simons Theory''},
\nbbstjournal{\doiref{10.1088/1126-6708/2008/10/053}{JHEP~0810,~053~(2008)}},
\nbbsteprint{\arxivref{0807.2063}{arxiv:0807.2063}}.

\bibitem{Aharony:2008ug}
\nbbstauthor{O.~Aharony, O.~Bergman, D.~L.~Jafferis and J.~Maldacena},
\nbbsttitle{``$\mathcal{N}$ = 6 superconformal Chern-Simons-matter theories,
  M2-branes and their gravity duals''},
\nbbstjournal{\doiref{10.1088/1126-6708/2008/10/091}{JHEP~0810,~091~(2008)}},
\nbbsteprint{\arxivref{0806.1218}{arxiv:0806.1218}}.

\bibitem{Beisert:2008cf}
\nbbstauthor{N.~Beisert and F.~Loebbert},
\nbbsttitle{``Open Perturbatively Long-Range Integrable gl(N) Spin Chains''},
\nbbstjournal{\doiref{10.1166/asl.2009.1034}{Adv.~Sci.~Lett.~2,~261~(2009)}},
\nbbsteprint{\arxivref{0805.3260}{arxiv:0805.3260}}.

\bibitem{Zwiebel:2008gr}
\nbbstauthor{B.~I.~Zwiebel},
\nbbsttitle{``Iterative Structure of the $\mathcal{N}$ = 4 SYM Spin Chain''},
\nbbstjournal{\doiref{10.1088/1126-6708/2008/07/114}{JHEP~0807,~114~(2008)}},
\nbbsteprint{\arxivref{0806.1786}{arxiv:0806.1786}}.

\end{thebibliography}
\ifarxiv
\bibliographystyle{nb}
\fi

\end{document}